\documentclass[12pt]{iopart}

\usepackage[cp866]{inputenc}
\usepackage[T2A]{fontenc}
\usepackage[english]{babel}

\usepackage{hyperref}
\usepackage{graphicx}
\begin{document}

\title[Potts spin glasses with 3, 4 and 5 states]{Potts spin glasses with 3, 4 and 5 states near $T=T_c$:\\
 expanding around the replica symmetric solution
}

\author{N.V. Gribova$^{1,2}$, T.I. Schelkacheva$^1$, E.E. Tareyeva$^1$}

\address{$^1$ Institute for High Pressure Physics, Russian Academy of
Sciences, Troitsk 142190, Moscow Region, Russia}
\address{$^2$ Institute for Computational Physics, University of
Stuttgart, Pfaffenwaldring 27, D-70569 Stuttgart, Germany}

\date{\today}

\begin{abstract}
Expansion for the free energy functionals of the Potts spin glass models
with 3, 4 and 5 states up to the fourth order in $\delta q_{\alpha \beta }$
around the replica symmetric solution (RS) is investigated using a
special quadrupole-like representation. The temperature dependence of the
1RSB order parameters is obtained in the vicinity of the point $T=T_c$
where the RS solution becomes unstable. The crossover from continuous to
jumpwise behavior with increasing of number of states is derived
analytically. The comparison is made of the free energy expansion for the
Potts spin glass with that for other models.
\end{abstract}
\submitto{\JPA}

\maketitle


The spin glass corresponding to the random variant of the Potts model
\cite{Wu} occured to be one of the first models where the scenario of the
transition to the nonergodic state was discovered \cite{ElSh83,GoEl85} to
be quite different from the standard one, that is of the
Sherrington--Kirkpatrick (SK) model \cite{sk}. Nevertheless, it remains to
be in the focus of investigations till now and a number of questions
still are not answered (see, e.g., the recent papers
Ref.\cite{Janis, Al} that are also dealing with the mean field Potts
glasses with few number of states).

It is well known that in the case of SK model the so callled full replica
symmetry breaking (FRSB)
\cite{Par} takes place at the instability point $T_c$ of RS solution. All
other models with two-particle interaction and with reflection symmetry
behave in the same way \cite{Full}.  The Potts glass models (with $p>2$)
are usually considered to be classical examples of models without
 reflection symmetry: FRSB does not work at $T_c$
\cite{ElSh83,GoEl85,Gross,NoSh93,Ritort,Gribova} and there is a region
where the 1RSB solution is stable (see, however,
Ref.\cite{Janis}). It is known (see, e.g., Ref.\cite{ Ritort}
and references therein) from the numerical solution, that if the number of
states $p\leq4$, then the 1RSB solution appears continuously at $T_c$ --
the bifurcation point of the RS solution. For $p>4$ the 1RSB solution
appears jumpwise. The Potts model with $p=4$ presents a boundary between
two types of phase transitions. In this case the order parameter $m$
defining the number of groups in 1RSB construction achieves its maximum
value $m=1$ just at the point $T_c=T_{1RSB}$. It is worthwhile to follow
in an analytic way, using the bifurcation theory, how the crossover from
continuous to jumpwise behavior of the order parameter happens. This paper
is devoted to this problem. To some extent we proceed in a same line as
the authors of the paper Ref.\cite{A.Crisantione} where the SK model
was considered.

It is worth to notice that the Potts symmetry leads to the cancellation of
some terms in the free energy. This by-turn
results in a trivial solution for the order parameters in  the RS
approximation. This is why the Potts models differ from other models
without reflection symmetry, in particular, from the models similar to SK
model, that is with the Hamiltonian
\begin{equation}
H=-\frac{1}{2}\sum_{i\neq j}J_{ij} U_i U_j, \label{SK}
\end{equation}
with quenched interactions $J_{ij}$  distributed with Gaussian
probability,
but without reflection symmetry:
\begin{equation}
Tr\left[{U}^{(2k+1)}\right]\neq 0 \label{odd}
\end{equation}
for some integer $k$. In this case the disorder smears out the
 phase transition: there is no trivial RS solution at high temperature
(see, e.g. \cite{Luchinskaya, Hydrogen}).

At the same time the Potts symmetry leads to the fact that the order
parameter is one and the same for all of $p$ states, that makes it related
to two--particle generalized SK--type models and allows to show that the
dependence of the free energy on the order parameters in the vicinity of
the bifurcation point $T_c$ is universal in both cases.

 We use a special representation of Potts models which originates from
 the analogy between three--state Potts spin glass
 and isotropic quadrupole glass (see below). In our representation the
 number of integrations is less than in usual approach and so it is more
 convenient for actual calculations. Moreover, this representation
 demonstrates explicitly the appearence of cubic terms in the free
 energy expansion. These terms play a crucial role for the
 order-parameter behavior.

The Potts glass model is defined by the Hamiltonian
\begin{equation}
\hat H=-\frac{p}{2}\sum_{i\neq j}J_{ij} {\delta_{\sigma_{i},\sigma_{j}}}.
\label{one1}
\end{equation}
where $p$ is the number of states and
variables $\sigma_i$, $\sigma_j$ can take the values $0,1,...,p-1$.
Here  $J_{ij}$ are random interactions
distributed with Gaussian probability
\begin{equation} P(J_{ij})=\frac{\sqrt{N}}{\sqrt{2\pi}
\tilde{J}}\exp\left[-\frac{(J_{ij})^{2}N}{ 2\tilde{J}^{2}}\right],
\label{two1}
\end{equation}
where the factor $N$ insures a sensible thermodynamic limit.

The 3-state Potts spin glass was described earlier in our papers
(see, e.g., Ref.\cite {L, Lu, Gribova}) using the representation in
terms of the operators of quadrupole momenta:
$Q=\left(3{J_{z}}^{2}-2\right)$,
$V=\sqrt{3}\left({J_{x}}^{2}-{J_{y}}^{2}\right)$, $\textbf{J}=1$,
$J_{z}=1,0,-1$. The operators $Q$ and $V$ commute and in the representation
where both of them are diagonal have the following form
$$
Q =
\left(
\begin{array}{ccc}
-2& 0& 0\\
\phantom{x}0& 1& 0\\
\phantom{x}0& 0& 1\\
\end{array}
\right)
\hspace{10mm}
V =
\left(
\begin{array}{cccc}
0& \phantom{x} 0& 0\\
0& \phantom{x} \sqrt{3}& 0\\
0& \phantom{x} 0& - \sqrt{3}
 \end{array}
\right)
$$
Throughout this paper we shall use the short-cut notations
 $Q_{k} =(-2,1,1)$ and $V_{k }=(0,\sqrt{3},-\sqrt{3})$.

The equation
\begin{equation}
\left[Q_{k
}Q_{l}+V_{k }V_{l }\right]+2=6\delta_{k ,l }.
\label{two13}
\end{equation}
provides the equivalence of the Potts model with $p=3$ to the model with
the Hamiltonian
\begin{equation}
 H=-\frac{1}{2}\sum_{i\neq j}\left[J_{ij}Q_{i}Q_{j}+G_{ij}{V}_{i}{V}_{j}\right] ,
\label{two834}
\end{equation}
(with $J_{ij}\equiv G_{ij}$).
The quenched interactions are distributed with Gaussian
probability.

Let us note that this representation is very useful for actual
calculations and allowed us to obtain, in particular, the number of
metastable states at zero temperature
\cite{Lu} and to determine the low--temperature boundary of the stability
of the 1RSB solution
\cite{Gribova}.

Using the replica method we obtain the averaged
free energy of the system in the form

\begin{eqnarray}
\label{two934}
 F/NkT=\lim_{n \rightarrow
0}\frac{1}{n}\left[-2t^{2}n+\frac{t^{2}}{2}\sum_{\alpha
\neq\beta}\left(q^{\alpha \beta }\right)^{2} \right.\nonumber
\\
\left.-\ln\Tr\exp\left[\frac{t^{2}}{2}\sum_{\alpha \neq\beta}q^{\alpha
 \beta }
\left({Q}^{\alpha}{Q}^{\beta}+{V}^{\alpha}{V}^{\beta}\right)\right]
\right],
\end{eqnarray}
where $q^{\alpha \beta }$ is the glass order parameter, $t={\tilde{J}}/kT$.
Here indices $\alpha$ and $\beta$  label replicas.
The Potts symmetry permits to describe the glass state with one
glass order parameter for both kinds of operators.

Using the standard procedure  we perform the first stage
of the replica symmetry breaking (1RSB) according to Parisi ($n$ replicas
are divided into  $n/m$ groups with  $m$ replicas in each group) and obtain
the free energy in the form (with $q^{\alpha \beta }= r_1$ if $\alpha $
and $\beta $ are from different groups and $q^{\alpha \beta }= r_1+v$ if
$\alpha $ and $\beta $ belong to the same group).

\begin{eqnarray}
\label{two137}
F_{\mathrm{1RSB}}=-NkT\times \nonumber
\\
\biggl\{t^2\left(2+(1-m)\frac{v^{2}}{2}-2v^{2}\right)+ \nonumber
\\
\frac{1}{m}\ln\int dz^G\int ds^G \left[\Psi\right]^{m}\biggr\}.
\end{eqnarray}

Here
\begin{eqnarray}
\Psi=\exp{\left(-2{\theta}_{1}\right)}+\exp{\left({\theta}_{1}\right)}\left[\exp{\left({\theta}_{2}\right)}
+\exp{\left(-{\theta}_{2}\right)}\right],\\
{\theta}_{1}=zt\sqrt{v},{\theta}_{2}=st\sqrt{3v},
\\
\int\frac{dz}{\sqrt{2\pi}}(\ldots)\exp\left(-\frac{z^2}{2}\right)\equiv \int dz^G(\ldots).
\end{eqnarray}

The Potts symmetry provides also the trivial solution
$q_{RS}=0$ and $r_{1}=0$ (see e.g.
Ref \cite{Ritort}).
Let us note that in general case of the Hamiltonian
(\ref{two834}) with $J_{ij}\neq G_{ij}$ one of the glass order parameter
can not be zero. This fact can be easily seen by analyzing the
high--temperature expansions for the RS equations for the order
parameters \cite{LuTa87}. Let us note that
the Eq.(\ref{two137}) can be obtained from the appropriate equation of
Ref. \cite{Ritort} by the integration over suitable linear
combinations of variables.

Expanding the free energy (\ref{two137}) near $T_c$ up
to $v^4$ and solving the equations that define the extremum
conditions for the free energy one can obtain~\cite{L}:
\begin{equation}
v = 8\tau -84\tau ^2,\hspace{3mm} m = \frac{1}{2} - \frac{9}{2} \tau,
\hspace{3mm}\Delta F= 16\tau ^3,
\label{mvf3}
\end{equation}
where $\Delta F=F_{1RSB}-F_{RS}$ and $ \tau = t-t_c$.

Let us consider now the case $p=4$. In this case one can use the following
operators $Q'\sim 3{J_{z}}^{2}-J(J+1)$, $V'\sim J_{z}$
for $\textbf{J}=3/2$
and the third operator $P'$ defined as to be orthogonal to $Q'$ and $V'$.
Namely, $2Q'_k= (1,-1,-1,1)$, $2\sqrt{5}V'_k=
(-3,1,-1,3)$, $2\sqrt{5}P'_k= (1,3,-3,-1).$

It is easy to see that the following equations hold:
$$Q'^2=\frac{1}{4};~~
V'^2=\frac{1}{4}+\frac{2}{5}Q';~~P'^2=\frac{1}{4}-\frac{2}{5}Q',$$
$$V'P'=-\frac{3}{10}Q';~~Q'V'= -\frac{3}{10}P'+\frac{2}{5} Q'; $$
$$Q'P'=\frac{2}{5}P'-\frac{3}{10}Q',$$
so that
\begin{equation}
Q'_kQ'_l+V'_kV'_l+P'_kP'_l +\frac{1}{4} = \delta _{k,l}.
\label{4444prs}
\end{equation}
There is no reflection symmetry and we have
\begin{equation}
\Tr P'^2Q'=-\frac{2}{5};\hspace{2mm}\Tr V'^2Q'=\frac{2}{5};\hspace{2mm}\Tr P'V'Q'=-\frac{3}{10},
\label{kuby}
\end{equation}
that leads to the appearance of cubic terms in the free energy
expansion\footnote{It is worth to notice that our representation in terms of three
matrices is not unique. For example, one can use the set: $A_k=(1,-1,-1,1)$,
$B_k=(-1,1,-1,1)$, $C_k= (1,1,-1,-1).$ We use, however, the
quadrupole--like operators in order to retain the physical meaning of
generalized anisotropic models and to preserve the analogy between even--p
and odd--p cases.}.

Now, using the replica method one can write the averaged free energy for
the Hamiltonian
\begin{equation}
 H=-\frac{1}{2}\sum_{i\neq
j}J_{ij}[Q'_{i}Q'_{j}+V'_{i}V'_{j}+P'_iP'_j] ,
\label{ham4}
\end{equation}
in the form analogous to (\ref{two934}) and perform the 1RSB.
The extremum conditions for this 1RSB free energy have the form of three
equations that look different only at the first glance.
In fact they are the
equations from Ref.\cite{Ritort}, integrated over different
variables and so have a common solution $v\equiv \langle \langle
Q'_\alpha Q'_\beta \rangle \rangle = \langle \langle V'_\alpha
V'_\beta \rangle \rangle = \langle \langle P'_\alpha P'_\beta
\rangle \rangle$. Expanding the 1RSB
free energy in the neighborhood of $t_c$ up to
$v^4$ we obtain (in terms of standard variables
(as, e.g., in \cite{Ritort})):
\begin{eqnarray}\label{df4}
\Delta F/Nkt= -\frac{1}{8}(-1+m)t^2 v^2 \times \nonumber
\\
\times [-6+6t^2+4(m-1)t^4v+(3m^2-3m-7)t^6v^2].
\end{eqnarray}
Here $t=t_c+\tau $, $m=m_0+\delta m$ and $\tau $, $v$ and $\delta m$ are small.
The extremum conditions relative to $v$ and $m$ give the system of
equations:
\begin{eqnarray}\label{min4}
-\frac{1}{2}(-1+m)t^2v\times \nonumber
\\
\times[-3 +3t^2+3(m-1)t^4v+(3m^2+3m-7)t^6v^2]=0;\nonumber
\\
\frac{1}{8}t^2v^2[6-6t^2-8(m-1)t^4v+(4+12m-9m^2)t^6v^2]=0.
\end{eqnarray}
To determine the relative orders of the small values
$\tau $, $v$ and $\delta m$, one has to use the so-called bifurcation
equation (see, e.g.
Ref.\cite{tren}) for the system (\ref{min4}). One can also check
directly that the equations
(\ref{min4}) become incompatible if
$v$ and $\delta m$  are supposed to be series in integer powers of
$\tau $. If $\delta
m=m_1 \sqrt{\tau }$ and $v=v_1\sqrt{\tau }$ (as it follows from the
bifurcation equation)
then we obtain $t_c=1$, $m_0=1$, and for $m_1$ and $v_1$ the
following system of equations can be obtained from (\ref{min4}):
$$6+3m_1v_1-7v_1^2=0;$$
$$12+8m_1v_1-7v_1^2=0.$$
The system has two solutions that differ only by the sign.
Taking into account that
$$\Delta F/NkT=\frac{1}{2}m_1^2v_1^3\tau ^{5/2},$$
and that on the physical branch $m<1$,
we choose
$$ v=\sqrt{12\tau /35}; ~~~m=1-\sqrt{21\tau /5}.$$
So, in the case of Potts spin glass with 4 states
$m=1$ at the point of bifurcation $t_c$, that is $t_{1RSB}=t_c$.

Let us consider now the Potts glass with
$p=5$ in an analogous way.
We shall use the following diagonal operators:
$P_k^{(1)}=(-4,1,1,1,1)$,
$P_k^{(2)}=(0,-3,1,-1,3)$, $P_k^{(3)}=\sqrt{5}(0,1,-1,-1,1)$,
$P_k^{(4)}=(0,1,3,-3,-1)$. The equivalence to the Potts model
follows from the relation:
$$P_k^{(1)}P_l^{(1)}+P_k^{(2)}P_l^{(2)}+P_k^{(3)}P_l^{(3)}+P_k^{(4)}P_l^{(4)}+4 =
20\delta _{k,l}.
$$
The nonzero cubic terms are
$${P^{(1)}}^3=-12;~~{P^{(2)}}^2P^{(1)}=4;~~{P^{(3)}}^2P^{(1)}=4;$$
 $${P^{(4)}}^2P^{(1)}=4;~~
{P^{(2)}}^2P^{(3)}=16/\sqrt{5};$$
$${P^{(4)}}^2P^{(3)}=-16/\sqrt{5};~~ P^{(1)}P^{(3)}P^{(4)}=-12/\sqrt{5}$$

Now the 1RSB free energy expansion near
$t_c$ has the following form (in standard variables):
\begin{eqnarray}\label{df5}
\Delta F/NkT= -\frac{1}{12}(-1+m)t^2 v^2 \times \nonumber
\\
\times [-12+12t^2+4(2m-1)t^4v+(6m^2+6m-35)t^6v^2].
\end{eqnarray}
The extremum conditions are
\begin{eqnarray}\label{min5}
-\frac{1}{3}(-1+m)t^2v\times \nonumber
\\
\times[-6 +6t^2+3(2m-1)t^4v+(6m^2+6m-35)t^6v^2]=0, \nonumber
\\
\frac{1}{12}t^2v^2[12-12t^2-4(4m-3)t^4v+(41-18m^2)t^6v^2]=0.
\end{eqnarray}

It is easy to obtain that now
$v$ and $\delta m$ can be presented as series in integer degrees of
$\tau $:
$v=v_1\tau +v_2 \tau ^2$, $m=m_0+m_1\tau $.
Here $v_1$, $v_2$, $m_0$ and $m_1$ satisfy the following
equations:
$$6-3v_1+4m_0v_1=0,$$
$$4-v_1+2m_0v_1=0$$
and
$$6v_2-12m_1-92=0,$$
$$6v_2-16m_1-43=0,$$
so that finally we obtain
$$t_c=1;\hspace{2mm}m=\frac{3}{2}+\frac{49}{4}\tau ;\hspace{2mm} v=-2\tau
+\frac{239}{6}\tau ^2$$
and the transition from RS to 1RSB
can not take place at the bifurcation point.

Here we would like to make a remark about the 5-state model without any
pretend to be rigorous. Basing on the fact that the numerically obtained
\cite{Ritort} value of $t_{1RSB}$ for $m=1$ is very close to our $t_c$, one
can hope to obtain the jump to the 1RSB solution already from the
equation (\ref{df5}) and to estimate qualitatively the characteristics in
question. In fact, if we expand (\ref{df5}) in the vicinity of $m=1$ and
let its first and second derivatives relative to $v$ be equal to zero,
we obtain $t_{1RSB}^2= 184/187$ and the jump $\Delta v = 561/8464$ (compare
with the table in \cite{Ritort}).

\begin{figure}[t]
\begin{center}
\includegraphics[width=8 cm]{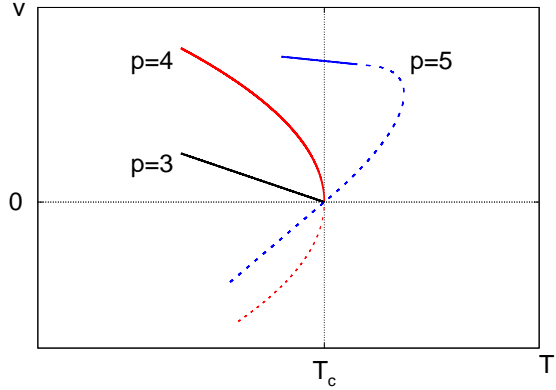}
\caption{(Colour online) Schematic representation of the order parameter $v$ in the vicinity of
the bifurcation point $t_c$ for $p=3,4,5$. Solid lines for $m\leq 1$,
dotted lines for $m>1$. }
\label{fig:scheme}
\end{center}
\end {figure}

Qualitative behavior of the solutions for $v$ near $t_c$ for $p=3,4,5$ is
presented on the figure \ref{fig:scheme}.

Now let us consider the 1RSB free energy expansion near the
bifurcation point $t_c$ from a more general point of view.
It is interesting
that the form of the series for $\Delta F$ in small
deviations $\delta q^{\alpha \beta}$ from $q_{RS}$ up to the third order
is one and the same for different models and coincides with the one that can be
written \cite{Hydrogen} for the random generalized Hamiltonian (\ref{SK})
where $U$ is arbitrary diagonal operator. The fact is that in the most
general case (including
(\ref{SK}) and (\ref{one1})) $\Delta F$ depends on the following (the
only nonzero) sums:
\begin{eqnarray}
&\lim_{n \rightarrow 0}\frac{1}{n}
{\sum_{\alpha,\beta}}^{'}(\delta
q^{\alpha\beta})^{2} =-[r-(m-1)v]^{2}-m(1-m)v^{2},\label{11eq:JP}\\
&\lim_{n \rightarrow 0}\frac{1}{n}
{\sum_{\alpha,\beta,\delta}}^{'}\delta
q^{\alpha\beta}\delta
q^{\alpha\delta} =[r-(m-1)v]^{2},
\label{51eq:JP}\\
&\lim_{n \rightarrow 0}\frac{1}{n}
{\sum_{\alpha,\beta}}^{'}(\delta
q^{\alpha\beta})^{3}=-[r-(m-1)v]^{3}+\nonumber\\
&3m(m-1)[r-(m-1)v]v^{2}+m(m-1)(2m-1)v^{3}, \label{61eq:JP}\\
&\lim_{n \rightarrow
0}\frac{1}{n}{\sum_{\alpha,\beta,\gamma}}^{'}\delta
q^{\alpha\beta}\delta q^{\beta\gamma}\delta
q^{\gamma\alpha}=2[r-(m-1)v]^{3} - \nonumber\\
&3m(m-1)[r-(m-1)v]v^{2}-m^{2}(m-1)v^{3}, \label{711eq:JP}\\
&\lim_{n \rightarrow
0}\frac{1}{n}{\sum_{\alpha,\beta,\gamma}}^{'}(\delta
q^{\alpha\beta})^{2}\delta
q^{\alpha\gamma}=[r-(m-1)v]^{3}- \nonumber
\\
&[r-(m-1)v]v^{2}, \label{71eq:JP}\\
&\lim_{n \rightarrow
0}\frac{1}{n}{\sum_{\alpha,\beta,\gamma,\delta}}^{'}\delta
q^{\alpha\beta}\delta q^{\alpha\gamma}\delta q^{\alpha\delta}=\nonumber\\
&\lim_{n \rightarrow
0}\frac{1}{n}{\sum_{\alpha,\beta,\gamma,\delta}}^{'}\delta
q^{\alpha\beta}\delta q^{\alpha\gamma}\delta q^{\beta\delta}=
-[r-(m-1)v]^{3},\label{81eq:JP}
\end{eqnarray}
with  $r=r_{1}-q_{RS}$.
The prime on the sum means that only the superscripts belonging to the same
 $\delta q$ are necessarily different in  $\sum'$ .

So, the deviation $\Delta F$ of the free energy $F_{1RSB}$ from its
replica symmetric part  in the most general case is

\begin{eqnarray}\label{1011frs}
\frac{\Delta F}{NkT}=\frac{t^2}{4}\left[1-t^{2}W\right]\left\{-\left[r-(m-1)v\right]^{2}-\right.
\nonumber\\
\left.v^{2}m(1-m)\right\}-\frac{t^{4}}{2}L\left[r-(m-1)v\right]^{2}- \nonumber\\
t^{6}\left\{C\left[r-(m-1)v\right]^{3}+\right. \nonumber
\\
D\left[r-(m-1)v\right]v^{2}m(m-1)-B_{3}v^{3}m^{2}(m-1)+\nonumber
\\
\left. B_{4}v^{3} m(m-1)(2m-1)\right\}+...
\end{eqnarray}
where
$t=t_c+\tau $, and  parameters
$W, L, C, D, B_3,
  B_4$ are some combinations of operators averaged over the RS solution.
 For example, the coefficient $L$ enters $\Delta F$ with the sum
 (\ref{51eq:JP}): $\frac{1}{n}
{\sum_{\alpha,\beta,\delta}}^{'}\delta q^{\alpha\beta}\delta
q^{\alpha\delta}$

Let us note that in the case of zero RS solutions for the order
parameters the expansion does not contain the terms where
some indices occur only once. In the case of reflection symmetry
there is no terms where some indices occur odd number of
times. If there is no reflection symmetry,
using
the extremum conditions for the free energy for Hamiltonian (\ref{one1})
and taking into account the fact that $L|_{t=t_c}\neq {0}$,  the
bifurcation condition gives\cite{Hydrogen}
$r-(m-1)v=0+o(\Delta t)^{2}$, i.e. the condition that there  is no linear
term for the glass order parameters. In fact, there is no other linear
term because
\begin{equation}
\left[1-t^{2}W\right]|_{t=t_c}=\lambda_{\rm RS repl}|_{t=t_c}=0
\label{40510prs}
\end{equation}
at the bifurcation point.

In the case of the Potts spin glass model the situation is quite
different, although the reflection symmetry is absent.
Now $L=0$ because it is zero RS solution that bifurcates
 so that single indices mean that $L$ is composed of zero values of
averaged
 $\Tr Q$,$\Tr V$.... Analogously the coefficients in front of
(\ref{71eq:JP}) and (\ref{81eq:JP}) are also zero. So, we have
$C=2B_{3}-B_{4}$ and $D=3(-B_{3}+B_{4})$. Here $B_{3}$ is the coefficient
in front of
(\ref{711eq:JP}), and $B_{4}$ that in front of (\ref{61eq:JP}).
Since $L|_{t=t_{0}}= {0}$, then the condition $r-(m-1)v=0$ is not fulfilled.
(If $p>2$ then $r_{1}=r=0$.\cite{Ritort})

In general case we obtain from the extremum conditions for the
free energy (\ref{1011frs}) the following equations for the order
parameters:
\begin{eqnarray}
& 2m(m-1)vZ\Delta t =
t_c^{6}m(m-1)v\left\{3\left[-B_{4}+\right.\right.\nonumber\\
&\left.\left. m(-B_{3} +2B_{4})\right]v+2D\left[r-(m-1)v\right]\right\} \label{30prs},\\
\nonumber \\
&(2m-1)v^{2}Z\Delta t =
t_c^{6}v^{2}\left\{(2m-1)\left[-B_{4}+\right.\right. \nonumber \\
&\left.m(-B_{3}+2B_{4})\right]v+m(m-1)(-B_{3}+2B_{4})v+\nonumber\\
&\left.D(2m-1)\left[r-(m-1)v\right]\right\}
\label{40prs},
\end{eqnarray}
where
\begin{equation}\label{B123eq:JP}
Z = \frac{d}{dt}{\left[\frac{t^2}{4}\left[1-t^{2}W\right]
\right]}_{\mid_{t_c}}.
\end{equation}

Hence we obtain from (\ref{30prs})-(\ref{40prs}):
\begin{equation}
m(m-1)\left[mB_{3}-B_{4}\right]=0.
\label{401prs}
\end{equation}

In fact, in the case of the Potts spin glass with $p=3$ for the Hamiltonian
 (\ref{two834})
at $t=t_{c}$ we have:
\begin{eqnarray}
W &= \frac{1}{2}\left[\left(TrQ^2/3\right)^{2}+
\left(TrV^2/3\right)^{2}\right]=4,\nonumber\\
B_{3}&=\frac{1}{6}\left[\left(TrQ^2/3\right)^{3}+
\left(TrV^2/3\right)^{3}\right]=\frac{8}{3},\nonumber\\
B_{4}&=\frac{1}{12}\left[\left(TrQ^3/3\right)^{2}+
3\left(TrQV^2/3\right)^{2}\right]=\frac{4}{3},\nonumber\\
D&=-4.
\label{0298prs}
\end{eqnarray}

In this case in the neighborhood of the bifurcation point one can get from the
equations (\ref{30prs})-(\ref{40prs}) that $v\sim \tau $,  and from the
 equation (\ref{401prs}) $m=\frac{1}{2}$.

Analogously, in the case of 4 states (for the Hamiltonian (\ref{ham4}) we
 obtain: $D=0$, $B_{3} = B_{4} = \frac{1}{128}$ so that $m=1$. The
 equation (\ref{30prs}) is the identity now. The r.h.s. of (\ref{40prs})
 becomes zero and one has to take into account higher order terms in the
 expansion of $F_{1RSB}$. This leads to $v\sim\sqrt{\tau }$. For $p=5$ we
 have $D=64, B_{3}=\frac{128}{3}, B_{4}=64$, so that $m=\frac{3}{2}$ and
 $v\sim -\tau $.

 It is worth noticing that from the free--energy expansion
 (\ref{1011frs}) it follows directly that the stability of the RS state
 is determined by the sign of
 $\lambda_{\rm RS repl}$, just as in the case of SK model
 \cite{Almeida}. To prove this fact let us consider small deviations from
 the RS solution:
$q^{\alpha  \beta}=q_{RS}+\delta q^{\alpha \beta}$
at arbitrary temperature and let us perform 1RSB. Using the notations
introduced before, (that is
  $\delta q^{\alpha \beta }= r$ if $\alpha $ and $\beta $ are
 from different groups and $\delta q^{\alpha \beta }= r+v$ if $\alpha $
and $\beta $ belong to the same group) we can write $F_{1RSB}-F_{RS}$
up to the second order in $\delta q$ in the form
(\ref{1011frs}) with arbitrary $t$. Since
$\left[1-t^{2}W\right]=\lambda_{\rm RS repl}$, it is easy to see
that for $m<1$ and $L \geq {0}$ the RS state is stable ($\Delta F<0$)
for the values of temperature such that
$\lambda_{\rm RS repl}>0$. The similar result is valid for all
subsequent stages of RSB because of
the Parisi rule
$$\lim_{n \rightarrow 0}\frac{1}{n}
{\sum_{\alpha,\beta}}^{'}(\delta
q^{\alpha\beta})^{2}<0$$

 To conclude, we introduced the representation for the Potts glass model
based on the operators of the quadrupole momenta. With the help of this
representation the expansions for the free energy functionals of the Potts spin glass models
with 3, 4 and 5 states up to the fourth order in $\delta q_{\alpha \beta }$
around the replica symmetric solution were obtained. The temperature
dependence of the 1RSB order parameters in the vicinity of the point
$T=T_c$ where the RS solution becomes unstable was derived. The crossover
from continuous to jumpwise behavior with the growing of the number of
states was traced analytically. The comparison was made of the free energy
expansions for the Potts spin glass with that for other models and the
similarity of such expansions was demonstrated.

Authors thank V.N. Ryzhov and N.M.Chtchelkatchev for
helpful discussions and valuable comments.

This work was supported in part by the Russian
Foundation for Basic Research (Grant No. 08-02-00781),
 and the Program of the Presidium of the Russian Academy of
Sciences.

\end{document}